\documentstyle[aps,tighten]{revtex}
\begin{document}
\preprint{EFEI-FIS}
\draft
\renewcommand{\thefootnote}{\fnsymbol{footnote}}
\title{The rotating detector and vacuum fluctuations}
\author{%
V. A. De Lorenci$^{1}$%
\protect\thanks{Electronic mail: \tt lorenci@cpd.efei.br}, 
R. D. M. De Paola$^{2}$%
\protect\thanks{Electronic mail: \tt rpaola@lafex.cbpf.br} 
and  N. F. Svaiter$^{2}$%
\protect\thanks{Electronic mail: \tt nfuxsvai@lafex.cbpf.br}
}
\address{$^{1}$
Instituto de Ci\^encias - Escola Federal de Engenharia de Itajub\'a \\
Av. BPS 1303 Pinheirinho, 37500-000 Itajub\'a, MG -- Brazil} 
\address{$^{2}$
Centro Brasileiro de Pesquisas F\'{\i}sicas\\
Rua Dr.\ Xavier Sigaud 150 Urca, 
22290-180 Rio de Janeiro, RJ -- Brazil} 

\date{\today}
\maketitle

\begin{abstract}
\hfill{\small\bf Abstract}\hfill\smallskip
\par
In this work we compare the quantization of a massless scalar field in 
an inertial frame with the quantization in a rotating frame. We used
the Trocheries-Takeno mapping to relate measurements in the inertial and the 
rotating frames. An exact solution of the  Klein-Gordon equation in the 
rotating coordinate system is found and the Bogolubov transformation
between the inertial and rotating modes is calculated, showing that the
rotating observer defines a vacuum state different from the Minkowski one. 
We also obtain the response function of an Unruh-De Witt detector coupled with 
the scalar field travelling in a uniformly rotating world-line. The response  
function is obtained for two different situations: when the quantum field is  
prepared in the usual Minkowski vacuum state and when it is prepared in the  
Trocheries-Takeno vacuum state. We also consider the case of an inertial 
detector interacting with the field in the rotating vacuum.
\end{abstract}
\renewcommand{\thefootnote}{\arabic{footnote}}

\section{Introduction}
The key point of special relativity is that the Poincar\'e group is the
symmetry group of all physical systems. The definitions of the Lorentz and
Poincar\'e groups are based as groups of mappings that leave invariant
the flat metric. The natural consequence is that an inertial observer 
(in an inertial reference frame) can assign a time and space location to
any event occurring in space-time, using light clocks, etc. To obtain a
``physical" interpretation of the Poincar\'e mapping we have to  derive
general relations between the space-time measurements made by different
observers who are in different inertial frames. 

So far, we have been considering only classes of measuring devices in inertial 
reference frames. However, suppose we are to make measurements with a device 
in non-inertial frames, as for example in a rotating disc. To discuss such
measurements and to show how they can be incorporated into a space-time 
description, entails that distance and time measurements made 
with some arbitrary set of measuring devices can always be made to correspond
to the coordinates of space-time by means of a suitable space-time mapping.
In other words, in order to compare measurements made by inertial and 
non-inertial (e.g. rotating) observers, we must present the mapping that 
relates the measurements made with the two different sets of devices.

For example, in a Galilean scenario it is possible to relate the space and
time measurements made in a rotating frame to those in an inertial one by
the mapping:
\begin{eqnarray}
\label{galileo1}
t&=&t',\\
r&=&r',\\
\theta&=&\theta'-\Omega t',\\
z&=&z',
\label{galileo2}
\end{eqnarray}
where $\Omega$ is the constant angular velocity around the $z$ axis of the
inertial frame. In the above, the cylindrical coordinate system 
$x'^{\mu}=\{t',r',\theta',z'\}$ is adapted to an inertial observer and the 
rotating coordinate system $x^{\mu}=\{t,r,\theta,z\}$ is the one adapted to 
the rotating observer. Although some authors tried to construct the counterpart 
of this mapping incorporating special relativity \cite{strauss}, the final 
answer for this question is still open. 

In two recent papers \cite{lorenci} it is assumed that the mapping which 
relates the inertial frame with the rotating one is given by:
\begin{eqnarray}
\label{tak1}
t &=&t'\cosh\Omega r' - r'\theta' \sinh\Omega r',\\
r &=&r',\\
\theta&=&\theta'\cosh\Omega r' - \frac{t'}{r'}\sinh\Omega r',\\
z&=&z'.
\label{tak2}
\end{eqnarray}
Such group of transformations was presented a long time ago
by Trocheries and also Takeno \cite{trocheries}. In Takeno's derivation,
three assumptions were made: (i) the transformation laws constitute a group;
(ii) for small velocities we must recover the usual linear velocity law 
($v=\Omega r$); and (iii) the velocity composition law is also in agreement 
with special relativity. In fact, the above transformation predicts that the
velocity of a point at distance $r$ from the axis is given by 
$v(r)=\tanh(\Omega r)$. 

It is our purpose, in this work, to investigate how does an uniformly rotating
observer see the Minkowski vacuum. In this direction we performed
the quantization of a scalar field as an observer who rotates uniformly around 
some fixed point would do it. We assume the Trocheries-Takeno
transformations (\ref{tak1}-\ref{tak2}) to compare measurements made in the 
inertial and in the rotating frames.

The canonical quantization of a scalar field in the rotating frame,
related to the inertial one through coordinate transformations 
(\ref{galileo1}-\ref{galileo2}), was made by Denardo and Percacci and also by
Letaw and Pfautsch \cite{letaw}. To compare the quantizations performed in the
inertial and rotating frames, the authors calculated the Bogolubov
transformation \cite{birrell} between the inertial modes 
$\psi_i(t',r',\theta',z')$ and the modes adapted to the rotating frame,
${\overline \psi}_j(t,r,\theta,z)$. Since they found that the Bogolubov
coefficients $\beta_{ij}$ are null, they conclude that the rotating vacuum or 
no-particle state, as defined by the rotating observer, is just the Minkowski
vacuum $\left|0,M\right>$.

Another way to compare different quantizations is to study the vacuum activity 
of a quantum field, and this is performed introducing a measuring device
which  couples with the quantum field via an interaction Lagrangian.
(In the following discussion we will take into account the detector model
due to Unruh and De Witt \cite{unruh,dewitt}.)
Using first-order perturbation theory, it is possible to calculate the
probability of excitation per unit proper time (excitation rate) of such
a detector, that is, the probability per unit time that the detector, 
travelling in a given world-line and initially prepared in its ground state,
to wind up in an excited state when it interacts with the field in a given
state \cite{nami}. As an example, consider that the detector is in an inertial 
frame and the field is prepared in the Minkowski vacuum state. In this 
situation the detector will remain in the ground state (null excitation rate). 
This is easily understood, because there are no inertial particles in the 
Minkowski vacuum state. On the other hand, if the detector is put in a 
world-line of an
observer with constant proper acceleration and the field is prepared again in 
the Minkowski vacuum, the detector
has a non-null probability to suffer a transition to an excited state. 
This is the so-called Unruh-Davies effect and the quantitative result is
in agreement with the fact that the Minkowski vacuum state is seen as a 
thermal state by the accelerated observer, with temperature proportional to 
its proper acceleration \cite{unruh}. The construction of a quantum field 
theory with the implementation of the Fock space in Rindler's manifold leads 
to define the Rindler vacuum state $\left|0,R\right>$. For completeness, in 
the situation when the field is prepared in such a state and the detector is 
uniformly accelerated, the detector remains inert. Again one is tempted to 
conclude that this is so because there are no Rindler particles in 
$\left|0,R\right>$ to be detected in the uniformly accelerated frame.

The agreement between the response of a detector and canonical quantum field
theory seems not to occur for more general situations. Indeed, as was shown by 
Letaw and Pfautsch, if the detector is put in a uniformly rotating world-line 
and the field is prepared in the Minkowski vacuum state $\left|0,M\right>$, it 
is found a non-null excitation rate, in spite of the fact that 
$\left|0,M\right>$ is considered as the vacuum state for a rotating observer, 
as discussed above. The rotating detector is excited even though there are 
no particles as an orbiting observer would define them (see also \cite{padmanabhan}).

Recently Davies {\it et al} \cite{davies} (see also \cite{levin})
solved this paradox, still assuming the Galilean coordinate 
transformations (\ref{galileo1}-\ref{galileo2}) between the inertial and 
rotating frames. First of all note that the world-line of an observer in
the rotating frame is an integral curve of the Killing vector
$\xi=(1-\Omega^2 r^2)^{-\frac1{2}}\partial / \partial t$, which is 
timelike only for $\Omega r<1$. Therefore, for a given angular velocity 
$\Omega$ there will be a maximum value of the radial coordinate 
$r_{max}=1 / \Omega$ (the light cylinder) for which an observer a distance 
$r>r_{max}$ will be moving faster than light. The Bogolubov coefficient 
$\beta$ is a scalar product over a spacelike hypersurface, where the radial 
coordinate ranges over $0\leq r<\infty$ and the notion of Bogolubov 
transformations outside the light cylinder becomes obscure. In order to 
circumvent this problem, Davies {\it et al} introduced a perfectly conducting 
cylinder with radius $a<r_{max}$ and they prove 
that the response of the detector vanishes when it is put a distance $r<a$ from 
the rotation axis. They conclude that ``a rotating particle detector
corotating  with a rotating vacuum state registers the absence of quanta",
although they continue to regard the Minkowski vacuum as the rotating vacuum
state.

The aim of this paper is two-fold. The first one is to present an exact solution
of the Klein-Gordon equation in the Trocheries-Takeno coordinate system and to
show that the Bogolubov coefficients $\beta$ between inertial and rotating
modes are not zero. This fact proves that there is a Trocheries-Takeno vacuum 
state adapted to rotating observers. The second one is to analyse the behavior 
of an apparatus device, a detector which is coupled with the scalar field
travelling in inertial or rotating world-lines, interacting with the field in
the Minkowski or the Trocheries-Takeno vacuum states.

We organize this paper as follows. In Section II we second quantize a massless
scalar field in Takeno's rotating coordinate system, and also compare this 
quantization with the usual one in the inertial frame via the calculation of 
the Bogolubov coefficients. In Section III we introduce the measuring 
apparatus -- the Unruh-De Witt detector \cite{unruh}. We 
calculate its response function when it is rotating and the field is prepared 
in two different states: the Trocheries-Takeno vacuum state and the Minkowski 
vacuum state. We also consider the case of an inertial detector interacting 
with the field in the rotating vacuum. Conclusions are given in Section IV. 
In this paper we use $\hbar=c=k_B=1$.

\section{Canonical quantization in the inertial and rotating frames}
In this section we will make a comparison between the quantizations of
the massless Klein-Gordon field performed in the inertial frame and in the 
rotating one (Trocheries-Takeno), when the two coordinate systems are related 
by the mapping (\ref{tak1}-\ref{tak2}). Such a comparison will be made by
calculating the Bogolubov transformation \cite{birrell} between the inertial
and rotating modes, solutions of the respective Klein-Gordon equations.
For the quantization in the inertial frame one chooses
cylindrical coordinates on $t'=$\,constant hypersurfaces and writes
the Klein-Gordon equation in terms of them. We just quote the results
of refs. \cite{letaw}. Positive-frequency modes (with respect to inertial
time $t'$), solutions of the Klein-Gordon equation, are found to be:
\begin{equation}
 v_{q'm'k'}(t',r',\theta',z')=N_1\,e^{ik'z'+im'\theta'}
 e^{-i\omega't'}J_{m'}(q'r'),
\label{mode}
\end{equation}
where $\omega'\,^2=q'\,^2+k'\,^2$, $J_{m'}(q'r')$ are Bessel functions 
well-behaved at the origin and $N_1=q'^{\frac1{2}}[2\pi(2\omega')^{1/2}]^{-1}$
is a normalization factor. In the above, $m'=0,\pm 1,\pm 2,\pm 3,\dots$,
$0\leq q'<\infty$ and $-\infty<k'<\infty$. In this way the field is expanded
as:
\begin{equation}
 \phi(t',r',\theta',z')=\sum_{m'}\int dq'\,dk'\,
 \left[b_{q'm'k'}v_{q'm'k'}(t',r',\theta',z')+
 b^{\dagger}_{q'm'k'}v^*_{q'm'k'}(t',r',\theta',z')\right],
\label{expansion}
\end{equation}
where the coefficients $b_{q'm'k'}$ and $b^{\dagger}_{q'm'k'}$ are, 
respectively, the annihilation and creation operators of the inertial
quanta of the field and satisfy the usual commutation rule 
$[b_i,b_j^{\dagger}]=\delta_{ij}$. In the above, the modes 
$v_i$ and $v^*_i$ are called,
respectively, positive and negative frequency modes with respect to
the Killing vector $\partial / \partial t'$. It is important to stress 
that in stationary geometries, such as the Minkowski space-time, the definition 
of positive and negative frequency modes has no ambiguities. The Minkowski 
vacuum state is then defined by
\begin{equation}
 b_{q'm'k'}\left|0,M\right>=0,\,\,\,\,\,\,\,\,\forall\,q',m',k'.
\label{MV}
\end{equation}

Now we shall consider the quantization in the rotating frame.
Assuming the mapping (\ref{tak1}-\ref{tak2}) to connect measurements made in
the rotating frame and those made in the inertial one, the line element 
in the rotating coordinates assumes the non-stationary form \cite{lorenci}:
\begin{equation}
 ds^2=dt^2-(1+P)dr^2-r^2d\theta^2-dz^2+2Qdrd\theta+2Sdtdr,
\label{lineel}
\end{equation}
where $P, Q$ and $S$ are given by:
\begin{eqnarray}
P&=&(\frac{Y}{r^2}+4\Omega\theta t)\sinh^2\Omega r-
\frac{\Omega}{r}(t^2+r^2\theta^2)\sinh2\Omega r+\Omega^2Y,\\
Q&=&r\theta \sinh^2\Omega r-\frac1{2}t \sinh2\Omega r+\Omega rt,\\
S&=&\frac{t}{r}\sinh^2\Omega r-\frac1{2}\theta \sinh2\Omega r-\Omega r\theta,
\label{pqs}
\end{eqnarray}
with $Y=(t^2-r^2\theta^2)$. Note that this metric presents no event 
horizons. In order to implement the canonical quantization
first we have to solve the Klein-Gordon equation in the Trocheries-Takeno
coordinate system:
\begin{equation}
 \Box \phi(t,r,\theta,z)=0.
\label{KG}
\end{equation}
It is possible to show that a complete set, basis in the space of solutions
of the Klein-Gordon equation is given by $\{u_{qmk},u^*_{qmk}\}$, in which
\begin{equation}
 u_{qmk}(t,r,\theta,z)=N_2\,e^{ikz}
 \exp\left[i\left(m\cosh\Omega r+\omega r\sinh\Omega r\right)\,\theta
 -i\left(\frac{m}{r}\sinh\Omega r+
\omega\cosh\Omega r\right)t\right]J_m(qr),
\label{mode2}
\end{equation}
where $\omega^2=q^2+k^2$ and $N_2$ is a 
normalization factor. Again, $m=0,\pm 1,\pm 2,\pm 3,\dots$,
$0\leq q<\infty$ and $-\infty<k<\infty$. One sees that these modes are
well-behaved throughout the whole manifold.
Making use of the transformations (\ref{tak1}-\ref{tak2}) one can show 
that these modes are of positive frequency by using the criterium of di Sessa
\cite{disessa}, which states that a given mode is of positive frequency if
it vanishes in the limit $(t')\rightarrow -i\infty$, where $t'$ is the
inertial time coordinate, while $u^*_j$ are modes of negative frequency.
In this way, the field operator is expanded in terms of these modes as:
\begin{equation}
 \phi(t,r,\theta,z)=\sum_{m}\int dq\,dk\,
 \left(a_{qmk}u_{qmk}(t,r,\theta,z)+
 a^{\dagger}_{qmk}u^*_{qmk}(t,r,\theta,z)\right),
\label{expansion2}
\end{equation}
where the coefficients $a_{qmk}$ and $a^{\dagger}_{qmk}$ are, respectively, 
the annihilation and creation operators of the Trocheries-Takeno 
quanta of the field. The vacuum state defined by the rotating observer is 
thus the Trocheries-Takeno vacuum state $\left|0,T\right>$ and it is given by
\begin{equation}
 a_{qmk}\left|0,T\right>=0,\,\,\,\,\,\,\,\,\forall\,q,m,k.
\label{TTV}
\end{equation}
The many-particle states of the theory can be obtained through successive
applications of the creation operators on the vacuum state.

We are now ready to compare both quantizations by using the Bogolubov
transformations \cite{birrell}. Since both sets of modes
are complete, one can expand modes (\ref{mode}) in terms of modes (\ref{mode2}) 
and vice-versa, the coefficients of this expansion being called Bogolubov
coefficients. For instance:
\begin{equation}
 u_i(x)=\sum_j \alpha_{ij}v_j(x)+\beta_{ij}v^*_j(x)
\label{Bog1}
\end{equation}
and conversely:
\begin{equation}
 v_j(x)=\sum_i \alpha^*_{ij}u_i(x)-\beta_{ij}u^*_i(x),
\label{Bog2}
\end{equation}
where $\alpha_{jj'}=\left(u_j,v_{j'}\right)$ and 
$\beta_{jj'}=-\left(u_j,v^*_{j'}\right)$, where the scalar product 
is defined by:
\begin{equation}
 \left(\phi,\chi\right)=-i\int_{\Sigma} d\Sigma^{\mu} \sqrt{-g}\,
 \left[\phi\,\partial_{\mu}\chi^*-\chi^*\partial_{\mu}\phi\right],
\label{scalarprod}
\end{equation}
with $d\Sigma^{\mu}=n^{\mu}d\Sigma$, where $n^{\mu}$ is a future-oriented
unit vector orthogonal to the spacelike hypersurface $\Sigma$. As 
$\Sigma$ we will choose the hypersurface $t'=0$, $t'$ being the inertial
time. The relevant coefficient for our present analysis is the $\beta$
coefficient since it is this coefficient which gives the content of
rotating particles in the Minkowski vacuum \cite{birrell}, and for such 
a calculation we need to express the rotating modes $u_j(x)$ in terms of 
the inertial coordinates, using transformations (\ref{tak1}-\ref{tak2}). 
In this way:
\begin{eqnarray}
 u_{qmk}(t',r',\theta',z')&=&N_2\,e^{ikz'}
 \exp\left(i\theta'\left[m\cosh 2\Omega r'+
 \omega r'\sinh 2\Omega r'\right]\right)\nonumber \\
 &\times& 
 \exp\left(-it'\left[\omega\cosh 2\Omega r'+
 \frac{m}{r'}\sinh 2\Omega r'\right]
\right)
J_m(qr')
\label{mode3}
\end{eqnarray}
and the Bogolubov coefficient $\beta_{jj'}$ is written as:
\begin{eqnarray}
 \beta_{jj'}&=&+i\int_{0}^{\infty}r'dr'\int_{-\infty}^{\infty}dz'
 \int_{0}^{2\pi}d\theta' \left[u_{j}(x')\frac{\partial v_{j'}(x')}{\partial t'}
 -v_{j'}(x')\frac{\partial u_j(x')}{\partial t'}\right]\nonumber \\
 &=&N_1N_2\int_{-\infty}^{\infty}dz'e^{i(k+k')z'}
 \int_{0}^{\infty}r'dr'
 \left[\omega'-\omega\cosh 2\Omega r'-
 \frac{m}{r'}\sinh 2\Omega r'\right]
 J_{m'}(q'r')J_m(qr')\nonumber \\
 &\times& 
 \int_{0}^{2\pi}d\theta'\exp
\left(i\theta'\left[m'+m\cosh 2\Omega r'+\omega r'\sinh 2\Omega r'
\right]\right).
\label{bog}
\end{eqnarray}
The first integral is easily evaluated to a delta function
$2\pi\delta(k+k')$, while the third one gives us:
$$
 \int_{0}^{2\pi}d\theta'\exp\left(i\theta'A_{m,m'}(r',\omega)\right)=
 \left(iA_{m,m'}(r',\omega)\right)^{-1}\left[\exp\left(2\pi
 iA_{m,m'}(r',\omega)\right)-1\right],
$$
where 
\begin{equation}
A_{m,m'}(r',\omega)=m'+m\cosh2\Omega r'+\omega r'\sinh 2\Omega
r'.
\end{equation}
Thus, we obtain
\begin{eqnarray}
 \beta_{jj'}&=&2\pi N_1N_2\,\delta(k+k')
 \int_{0}^{\infty}r'dr'
 \left(\omega'-\omega\cosh 2\Omega r'-
\frac{m}{r'}\sinh 2\Omega r'\right)
 J_{m'}(q'r')J_m(qr')\nonumber \\
 &\times&
 \left(iA_{m,m'}(r',\omega)\right)^{-1}\left[\exp\left(2\pi
 iA_{m,m'}(r',\omega)\right)-1\right].
\label{bog2}
\end{eqnarray}

The resulting expression is difficult to evaluate, but nonetheless it is 
non-zero. (In the appendix we give an indirect proof that it is non-zero.)
This means that the two vacua considered are non-equivalent,
i.e., $\left|0,M\right>\neq\left|0,T\right>$, which means that the Minkowski
vacuum $\left|0,M\right>$ contains rotating quanta, i.e., Trocheries-Takeno
particles \cite{birrell}.

\section{Detector excitation rate}
We now pass to consider the probability of excitation of a detector which 
is moving in a circular path at constant angular velocity $\Omega$ and at
a distance $R_0$ from the rotation axis, interacting with the scalar field. 
The initial state of the detector is its ground state and for the initial 
state of the field we will consider the two vacuum states: the usual Minkowski 
vacuum state and also the Trocheries-Takeno vacuum state. The interaction 
with the field may cause transitions between the energy levels of the detector 
and if it is found, after the interaction, in an excited state, one can say 
that it has detected a vacuum fluctuation of the field \cite{levin}.

As a detector we shall be considering mainly the detector model of 
Unruh-De Witt \cite{unruh,dewitt}, which is a system with two internal 
energy eigenstates with monopole matrix element between these two states 
different from zero. According to standard theory \cite{nami}, the probability 
of excitation per unit proper time of such a system (normalized by the 
selectivity of the detector), or simply, its excitation rate, is 
given by:
\begin{equation}
 R(E)=\int_{-\infty}^{\infty} d\Delta t\,e^{-iE\Delta t}G^{+}(x,x'),
\label{rate}
\end{equation}
where $\Delta t=t-t'$, $E>0$ is the difference between the excited and ground 
state energies of the detector and $G^{+}(x,x')$ is the positive-frequency
Wightman function calculated along the detector's trajectory. Let us note that 
the positive-frequency Wightman function is given by
\begin{equation}
 G^{+}(x,x')=\left<0|\phi(x)\phi(x')|0\right>,
\label{wightman}
\end{equation}
where $\left|0\right>$ is the vacuum state of the field, which can either be
$\left|0,M\right>$ or $\left|0,T\right>$. Let us consider first the second
possibility.

If one splits the field operator in its positive and negative frequency parts
 with respect to the Trocheries-Takeno time coordinate $t$,
as $\phi(x)=\phi^{+}(x)+\phi^{-}(x)$, where 
$\phi^{+}(x)$ contains only annihilation operators and $\phi^{-}(x)$ contains 
only creation operators (see Eq.(\ref{expansion2})),
and also considers $\left|0\right>$ as the Trocheries-Takeno vacuum state,
i.e., $\left|0\right>=\left|0,T\right>$ then, using Eq.(\ref{TTV}), one
finds that:
\begin{equation}
 G_{T}^{+}(x,y)=\sum_iu_i(x)u^*_i(y),
\label{wightman2}
\end{equation}
where the subscript $T$ stands for the Wightman function calculated in the
Trocheries-Takeno vacuum state. Considering now the modes given by Eq.(\ref{mode2}) 
and that we are interested
in the situation where the detector is at rest in the Trocheries-Takeno frame,
i.e., $\theta=$\,constant, $z=$\,constant and $r=R_0=$\,constant, one finds:
\begin{equation}
 G_{T}^{+}(x,y)=\sum_{m=-\infty}^{\infty}\int_{0}^{\infty}dq
 \int_{-\infty}^{\infty}dk\,N_2^2\,
 e^{-i[\frac{m}{R_0}\sinh\Omega R_0+\omega\cosh\Omega R_0]\Delta t}J_m^2(qR_0).
\label{wightman3}
\end{equation}
Putting the above expression in Eq.(\ref{rate}), we find:
\begin{equation}
 R_T^{(r)}(E,R_0)=\sum_{m=-\infty}^{\infty}\int_{0}^{\infty}dq
 \int_{-\infty}^{\infty}dk\,N_2^2\,J_m^2(qR_0)
 \int_{-\infty}^{\infty}d\Delta t\,
 e^{-i[E+\frac{m}{R_0}\sinh\Omega R_0+\omega\cosh\Omega R_0]\Delta t}.
\label{rate2}
\end{equation}
(In the above, the subscript $T$ stands for the Takeno vacuum and the
superscript $(r)$ stands for the rotating world-line followed by the detector.)
The last integral gives us 
$2\pi\delta\left(E+\frac{m}{R_0}\sinh\Omega R_0+\omega\cosh\Omega R_0\right)$,
for which the argument is non-null only if $m<0$; we can take the summation
index to run for $m=1,2,3,...$, leaving us with 
\begin{equation}
 R_T^{(r)}(E,R_0)=2\pi\sum_{m=1}^{\infty}\int_{0}^{\infty}dq
 \int_{-\infty}^{\infty}dk\,N_2^2\,J_m^2(qR_0)\,
 \delta\left(E-\frac{m}{R_0}\sinh\Omega R_0
 +\omega\cosh\Omega R_0\right).
\label{rate3}
\end{equation}
The above expression predicts excitation for the detector, and depends in a
non-trivial way on the position $R_0$ where it is put. So we once again 
arrive at the same confrontation between canonical quantum field theory and 
the detector formalism, which was settled by Letaw and Pfautsch and 
Padmanabhan: how is it possible for the {\it orbiting} detector to be excited 
in the {\it rotating} vacuum? However a crucial distinction exists between our 
present analysis and the above-mentioned works: we state, as proved in the 
last section, that the rotating vacuum {\bf is not} the Minkowski vacuum.
We now analyse the two independent origins of the non-null excitation
rate, Eq.(\ref{rate3}).

Note that the present situation of an Unruh-De Witt detector being 
excited when put in an orbiting world-line interacting with the field in the
rotating (Trocheries-Takeno) vacuum is to be contrasted with the two following
situations. In fact, this same detector {\it is not excited} whether it is in
an inertial world-line and interacting with the field in the inertial 
(Minkowski) vacuum \cite{birrell,nami} or when it is uniformly 
accelerated and interacting with the field in the accelerated (Rindler) vacuum 
\cite{ginzburg}. However note that both Minkowski and Rindler space-times
are static, differently of the Trocheries-Takeno metric, and differently
also of the rotating metric obtained using the galilean transformation, which
are examples of non-static metrics. Recall that using instead the galilean 
transformation to a rotating frame it was also found by Letaw and Pfautsch a 
non-null excitation rate for the orbiting detector in the rotating vacuum, 
considered by them as $\left|0,M\right>$. Therefore the excitation in the 
present case may be attributed to the non-staticity of the Trocheries-Takeno 
metric.

The other origin of the excitation found above for the detector
can be atributed to the detector model we adopted \cite{ford}. Indeed, note 
that splitting the field operator in its positive and negative frequency 
parts with respect to rotating time $t$ in Eq.(\ref{wightman}), one can 
express the Wightman function as:
\begin{eqnarray}
 G^{+}(x,x')&=&\left<0|\phi(x)\phi(x')|0\right>\nonumber \\
 &=&\left<0|\phi^{+}(x)\phi^{+}(x')|0\right>+
 \left<0|\phi^{+}(x)\phi^{-}(x')|0\right>\nonumber \\
 &+&\left<0|\phi^{-}(x)\phi^{+}(x')|0\right>+
 \left<0|\phi^{-}(x)\phi^{-}(x')|0\right>.
\label{wightman4}
\end{eqnarray}
In the case where $\left|0\right>=\left|0,T\right>$, because of
Eq.(\ref{TTV}) only the second term above is non-vanishing, corresponding
to the emission (creation) of a Trocheries-Takeno quantum with simultaneous 
excitation of the detector and this is the term responsible for the 
non-vanishing excitation rate, Eq.(\ref{rate3}). In the context of quantum 
optics photodetection
is regarded as photo{\it absorption} processes only \cite{glauber,moyses}, a 
detector being able to be excited only when it absorbs 
(annihilates) a quantum of the field. In this way terms like the second one 
above are discarded and only terms like the third one are taken into account, 
such a procedure being called the rotating-wave approximation \cite{moyses}. 
Therefore purely absorptive detectors (Glauber model) always give vanishing
excitation rate in the vacuum state of the field. From this discussion, it is 
clear that the Glauber detector model will not be excited when put in the 
orbiting world-line and the field is in the Trocheries-Takeno vacuum state. 
Another context in which the inclusion of the antiresonant
term (second one above) plays a crucial role is that of accelerated observers,
where the thermal character of the Minkowski vacuum as seen by a Rindler
observer is not revealed if one uses the Glauber correlation function, but
only if one uses the Wightman one. In effect, the Wightman 
correlation function includes the vacuum fluctuations that are omitted in the 
Glauber function \cite{ford}, and to these very vacuum fluctuations can be 
attributed the non-vanishing excitation rate Eq.(\ref{rate3}). Because of
this feature the model of Unruh-De Witt is also called a fluctuometer 
\cite{sciama}.

We now discuss the other case of putting the detector
in an orbiting trajectory and preparing the scalar field in the usual inertial 
vacuum $\left|0,M\right>$. Writing $\left|0,M\right>$ for $\left|0\right>$
in Eq.(\ref{wightman}), it is easy to show that the positive frequency
Wightman function is given by:
\begin{equation}
 G_{M}^{+}(x',y')=\sum_j v_j(x') v^*_j(y'),
\label{wightmanM}
\end{equation}
where $M$ stands for the Minkowski vacuum state. As the rate of excitation
Eq.(\ref{rate}) is given in terms of the detector's proper time, we shall
express Eq.(\ref{wightmanM}) in terms of the rotating coordinates, using the 
inverse of Takeno's transformations. Let us begin with 
$G^{+}_M(x',y')$, in inertial coordinates, with identifications
$r'_1=r'_2=R_0$ and $z'_1=z'_2$, as demanded for this case:
\begin{equation}
 G^{+}_M(x',y')=\sum_{m=-\infty}^{\infty}\int_{0}^{\infty}dq
 \int_{-\infty}^{\infty}dk N_1^2 
 e^{-i\omega (t'_1-t'_2)+im(\theta'_1-\theta'_2)}J_m^2(qR_0).
\label{inert}
\end{equation}
The inverse of Takeno's transformations read
\begin{eqnarray}
\label{tak1b}
t' &=&t\cosh\Omega r + r\theta \sinh\Omega r,\\
r' &=&r,\\
\theta'&=&\theta\cosh\Omega r + \frac{t}{r}\sinh\Omega r,\\
z' &=&z.
\label{tak2b}
\end{eqnarray}
Using the above in Eq.(\ref{inert}) and taking note of the fact that the
detector is at rest in the rotating frame, i.e., $\theta_1=\theta_2$, we
see that in this manner the Minkowski Wightman function is a function of
the difference in proper time $\Delta t=t_1 - t_2$, which allows us to
calculate the rate of excitation of the orbiting detector when the field 
is in the Minkowski vacuum:
\begin{equation}
 R_M^{(r)}(E,R_0)=2\pi\sum_{m=1}^{\infty}\int_{0}^{\infty}dq
 \int_{-\infty}^{\infty}dk\,N_1^2\,J_m^2(qR_0)\,
 \delta\left(E-\frac{m}{R_0}\sinh\Omega R_0
 +\omega\cosh\Omega R_0\right).
\label{rateM}
\end{equation}
The result above is very much like Eq.(\ref{rate3}), with the exception
that in the above it appears the normalization of the inertial modes
$N_1$ instead of $N_2$.

Finally, let us suppose that it is possible to prepare the field in the
rotating vacuum and the detector is in an inertial world-line and let
us calculate the excitation rate in this situation:
\begin{equation}
 R_T^{(i)}(E,R_0)=\int_{-\infty}^{\infty} d\Delta t'\,
 e^{-iE\Delta t'}G_T^{+}(x',y'),
\label{rate5}
\end{equation}
where the superscript $(i)$ stands for the inertial world-line followed by
the detector, $\Delta t'$ is the difference in proper time in the inertial
frame, and $G_T^{+}(x',y')$ is given by Eq.(\ref{wightman2}), but written
now in terms of the inertial coordinates. It is 
not difficult to write $G_T^{+}(x',y')$ in terms of the inertial
coordinates, recalling that now the detector is not at rest in the rotating
frame. We have therefore the result that:
\begin{eqnarray}
 R_T^{(i)}(E,R_0)&=&2\pi\sum_{m=-\infty}^{\infty}\int_{0}^{\infty}dq
 \int_{-\infty}^{\infty}dk\,N_2^2\,J_m^2(qR_0)\nonumber \\
 &\times&
 \delta\left(E-\left(\omega\Omega R_0-\frac{m}{R_0}\right)\sinh(2\Omega R_0)
 -(m\Omega-\omega)\cosh(2\Omega R_0)\right).
\label{rate6}
\end{eqnarray}
In order to study the activity of the Trocheries-Takeno vacuum, we calculated
the rate of excitation of an Unruh-De Witt detector in two different 
situations: 
when it is put in the orbiting and in the inertial world-lines. Since in the
first case we found a non-null rate, contrary to the idea that the orbiting
detector co-rotating with the rotating vacuum should not perceive anything,
it can be considered as a noise of the rotating vacuum, being it perceived
regardless of the state of motion of the detector. This amounts to say that
the inertial detector will also measure this noise, and we normalize the rate
in this situation by subtracting from it the value Eq.(\ref{rate3}), resulting
in a normalized excitation rate for the inertial detector in interaction with
the field in the rotating vacuum:
\begin{eqnarray}
 &&{\cal R}_T^{(i)}(E,R_0)=2\pi\sum_{m=-\infty}^{\infty}\int_{0}^{\infty}dq
 \int_{-\infty}^{\infty}dk\,N_2^2\,J_m^2(qR_0)\times\nonumber \\
 &&\left[\delta\left(E-\left(\omega\Omega R_0-\frac{m}{R_0}\right)
 \sinh(2\Omega  R_0)-(m\Omega-\omega)\cosh(2\Omega R_0)\right)
 -\delta\left(E-\frac{m}{R_0}\sinh\Omega R_0
 +\omega\cosh\Omega R_0\right)\right].
\label{rateT2}
\end{eqnarray}
It remains to be clarified the meaning of $R_0$ in the excitation above. When
studying the quantization in the rotating frame, one has to choose the 
world-line followed by the rotating observer, and this is parametrized by two
quantities: the angular velocity $\Omega$ and the distance $R_0$ from the
rotation axis. The vacuum state which appears in such a quantization is thus
also indexed by these parameters and this is the origin of $R_0$ in the above.
A similar dependence of the excitation rate of a detector on a geometrical
parameter appears, for instance, in the well-known Unruh-Davies effect: a
uniformly accelerated detector interacting with the field in the Minkowski 
vacuum state will absorb particles in the same way as if it were inertial and
interacting with the field in a thermal bath, with a temperature that depends
on the proper acceleration of the detector.

\section{Summary and discussions}
In this work we quantize a massless scalar field in a uniformly rotating
frame and compare this quantization with the usual one in an inertial frame.
As a difference with regard to previous works, we assume a coordinate 
transformation between both frames that takes into account the finite velocity
of light and is valid in the whole manifold. In doing so, a material point 
orbiting around the axis of rotation never exceeds the speed of light, no 
matter how far it is from the axis. The metric, when written in rotating 
coordinates, presents no event horizons, although it is non-static and 
non-stationary. We recourse to a criterium of di Sessa to define positive and 
negative frequency modes in the rotating frame and the field is quantized along 
these lines. Such a quantization entails a vacuum state and by using the
Bogolubov transformations we were able to show that this vacuum state is 
inequivalent to the Minkowski one. This is the main result of the paper.
This results in that the Minkowski vacuum is seen as a many-rotating-particle 
state by a uniformly rotating observer, although it can not be seen as a 
thermal state (of Takeno particles), as in the case of a uniformly accelerated 
observer (for Rindler particles) and thus we cannot assign a temperature to it. 

We obtain the response function of an Unruh-De Witt detector in three different 
situations: the orbiting detector is interacting with the field prepared in the 
rotating and in the Minkowski vacuum states, and finally the detector is 
travelling in an inertial world-line and the field is prepared in the rotating 
vacuum. In the first case it is found that the detector gets excited and we 
attributed this excitation to two different causes: Firstly, we are using the 
Unruh-De Witt detector model instead of a purely absorptive detector as the 
Glauber's model, and secondly the Trocheries-Takeno metric is non-static. 
Because the rotating vacuum excites even a rotating detector, we consider this 
as a noise which will be measured by any other state of motion of the detector. 
In this way, when calculating the reponse of the inertial detector in the 
presence of the rotating vacuum we subtract from it this noise.

\acknowledgements
This work was partially suported by {\em Conselho Nacional de
Desenvolvimento Cient\'{\i}fico e Tecnol\'ogico} (CNPq)
of Brazil.

\section{Appendix}
What interests us here is the number of Trocheries-Takeno particles in a
given state $j=(q,m,k)$ that is present in the Minkowski vacuum, given in 
terms of the Bogolubov coefficients \cite{birrell}:
\begin{equation}
 \left<0,M\right|a_j^{\dagger}a_j\left|0,M\right>=
 \sum_{j'}\left|\beta_{jj'}\right|^2.
\end{equation}
We see that this is given by a sum of the squared modulus of the various 
coefficients. If we can show that at least one of these coefficients is
non-null, so we prove that there is a non-vanishing content of 
Trocheries-Takeno particles in the Minkowski vacuum.

We now give an indirect proof that the Bogolubov coefficient $\beta$ is in
fact non-zero. The calculations in the text show that:
\begin{eqnarray}
 \beta_{jj'}&=&2\pi N_1N_2\,\delta(k+k')
 \int_{0}^{\infty}r'dr'
 \left(\omega'-\omega\cosh 2\Omega r'-
\frac{m}{r'}\sinh 2\Omega r'\right)
 J_{m'}(q'r')J_m(qr')\nonumber \\
 &\times&
 \left(iA_{m,m'}(r',\omega)\right)^{-1}\left[\exp\left(2\pi
 iA_{m,m'}(r',\omega)\right)-1\right],
\label{Bog}
\end{eqnarray}
where 
\begin{equation}
A_{m,m'}(r,\omega)=m'+m\cosh2\Omega r'+\omega r'\sinh 2\Omega
r'.
\end{equation}

Let us study the high-velocity limit, $\Omega\rightarrow\infty$,
of $\beta_{00}$, which is for $m=0$ and $m'=0$:
\begin{equation}
 \beta_{00}=-2\pi i N_1(m'=0)N_2(m=0)\delta(k+k')
 \int_0^{\infty}rdr \frac{\omega'-\omega \cosh(2\Omega r)}
 {\omega r \sinh(2\Omega r)}J_0(qr)J_0(q'r)
 \left[e^{2\pi i\omega r \sinh(2\Omega r)}-1\right].
\label{00}
\end{equation}
As $\frac1{\sinh x}\rightarrow 0$ and $\frac{\cosh x}{\sinh x}\rightarrow 1$ 
in the limit $x\rightarrow\infty$, we find:
\begin{equation}
 \lim_{\Omega\rightarrow\infty}\beta_{00}=-2\pi i N_1(m'=0)N_2(m=0)\delta(k+k')
 \int_0^{\infty}dr J_0(qr)J_0(q'r)
 \left[1-e^{2\pi i\omega r \sinh(2\Omega r)}\right].
\label{limit00}
\end{equation}
Calling $K(q,q',\omega,\Omega)$ the integral above, note that:
\begin{equation}
 K(q,q',\omega,\Omega)=\Re \left\{K(q,q',\omega,\Omega)\right\} +
                       i\Im \left\{K(q,q',\omega,\Omega)\right\},
\label{K}
\end{equation}
where the real and imaginary parts read:
\begin{equation}
 \Re \left\{K(q,q',\omega,\Omega)\right\}=\int_0^{\infty}dr J_0(qr)J_0(q'r)
 \left(1-\cos\left[2\pi\omega r \sinh(2\Omega r)\right]\right)
\label{real}
\end{equation}
and
\begin{equation}
 \Im \left\{K(q,q',\omega,\Omega)\right\}=\int_0^{\infty}dr J_0(qr)J_0(q'r)
 \sin\left[2\pi\omega r \sinh(2\Omega r)\right].
\label{imag}
\end{equation}
Note that in Eq.(\ref{real}) above, the second integral is not capable to
make $\Re \left\{K(q,q',\omega,\Omega)\right\}$ vanish if the first one is
non-zero. And note that this is indeed the case, since one can find the first 
of them in Gradshteyn \cite{gradshteyn}, in terms of hypergeometric functions:
\begin{equation}
 \int_0^{\infty}dr J_0(qr)J_0(q'r)=\frac1{q+q'}
 F\left[\frac1{2};\frac1{2};1;\frac{4qq'}{(q+q')^2}\right],
\label{hypergeom}
\end{equation}
and it is non-zero.

We now present a different route to prove that $\beta$ is different from 
zero. We start from Eq.(\ref{Bog}); as for $m=m'=0$ the integrands go like
$\frac1{\sinh(2\Omega r)}$, and
$\frac1{\sinh(2\Omega r)}$ diverges for $r\rightarrow 0$, let us calculate
$\beta_{0,1}$, which is for $m=0$ and $m'=1$:
\begin{eqnarray}
 \beta_{01}&=&-2\pi i N_1(m'=1)N_2(m=0)\,\delta(k+k')
 \int_{0}^{\infty}rdr
 \frac{\left(\omega'-\omega\cosh 2\Omega r\right)}
 {\left(1+r\omega\sinh 2\Omega r\right)}
 J_{0}(qr)J_1(q'r)\nonumber \\
 &\times&
 \left[\exp\left(2\pi
 i(1+r\omega\sinh 2\Omega r)\right)-1\right].
\label{bog3}
\end{eqnarray}
So we call $I$ the integral above:
\begin{equation}
 I=\int_{0}^{\infty}rdr
 \frac{\left(\omega'-\omega\cosh 2\Omega r\right)}
 {\left(1+r\omega\sinh 2\Omega r\right)}
 J_{0}(qr)J_1(q'r)
 \left[\exp\left(2\pi
 i(1+r\omega\sinh 2\Omega r)\right)-1\right].
\label{I2}
\end{equation}
The first integral is the only one which is $\omega'-$dependent, i.e., 
it is a function of $\omega'$. So, if we prove that it is different from
zero, we prove that $\beta_{01}\neq 0$, since the second integral is not
sufficient to make it zero. Let us call it $I(\omega')$:
\begin{equation}
 I(\omega')=\omega'\int_{0}^{\infty}rdr
 \frac{J_{0}(qr)J_1(q'r)}
 {\left(1+r\omega\sinh 2\Omega r\right)}
 \left[\exp\left(2\pi ir\omega\sinh 2\Omega r\right)-1\right].
\label{I3}
\end{equation}
It is possible to separate $I(\omega')$ in its real and imaginary parts and,
using the same reasoning as for Eq.(\ref{real}), we can concentrate in the
second integral above, which we call $I_1(\omega')$:
\begin{equation}
 I_1(\omega')=-\omega'\int_{0}^{\infty}rdr
 \frac{J_{0}(qr)J_1(q'r)}
 {\left(1+r\omega\sinh 2\Omega r\right)}.
\label{I4}
\end{equation}
There is not an analytical expression for $I_1(\omega')$, but using the
Maple one can calculate particular values, such as:
\begin{equation}
 \int_{0}^{\infty}rdr
 \frac{J_{0}(r)J_1(r)}
 {\left(1+r\sinh r\right)}=0.183096;
\label{I5}
\end{equation}
so we proved, very indirectly, that $\beta\neq 0$.

\end{document}